\documentclass{article}
\usepackage[totalwidth=480pt, totalheight=680pt]{geometry}

\newtheorem{theorem}{Theorem}[section]

\newtheorem{proposition}[theorem]{Proposition}

\newcommand{\E}{{\rm E}}
\newcommand{\Var}{{\rm Var}}
\newcommand{\Covar}{{\rm Covar}}

\begin{document}

\title{Calibration of One- and Two-Factor Models For Valuation of Energy Multi-Asset Derivative Contracts}

\author{Josh Gray\\ 
\texttt{Josh.Gray@msci.com}\\
\\
Konstantin Palamarchuk \footnote{Both authors are at MSCI Energy Research Department, FEA group} \\
\texttt{Konstantin.Palamarchuk@msci.com}}

\maketitle
\begin{abstract}
We study historical calibration of one- and two-factor models that are known to describe relatively well the dynamics of energy underlyings such as spot and index natural gas or oil prices at different physical locations or regional power prices. We take into account uneven frequency of data due to weekends, holidays, and possible missing data. We study the case when several one- and two-factor models are used in the joint model with correlated model factors and present examples of joint calibration for daily natural gas prices at several locations in the US and for regional hourly power prices.
\end{abstract}

\newpage

\tableofcontents

\section{Overview}

As is well-known the calibration of valuation models is a crucial step towards the realistic valuation of financial contracts. In this paper we focus on historical calibration of one- and two-factor models that describe the dynamics of energy underlyings such as spot prices of natural gas at a physical location or regional market power prices. These models are known to perform relatively well for energy underlyings, for example, see \cite{Eydel} or \cite{FEAEnergy}. From these simple models the joint model of simultaneous evolution of several underlyings can be constructed by specifying the term structure of correlations between model factors. This joint model can be used for valuation of energy derivative contracts that depend on several underlyings, such as swing, transport, or structured contracts that are abundant in the energy OTC markets. These models are extensively used and tested by various FEA clients for valuation and risk management of OTC and standard energy contracts.

We choose a simple robust approach for the joint calibration of these models. First, we calibrate each model separately based on the historical time series of prices. After that, using the calibrated parameters and the time series of prices, we find the time series of stochastic factors of these models and compute the correlation between these factors to obtain the correlation term structure of the joint model. 

As the result of calibration, we obtain volatility and correlation term structures and mean-reversion rates together with confidence intervals for these estimates. We also obtain historical mean-reversion level term structures, which include the market price of risk and are less relevant for valuation since it is done in the risk-neutral framework.
After calibration is performed, we run the statistics analysis of model factors time series to see if they satisfy the initial modelling assumptions and if the model is a good fit for the data.

One of our goals is providing rigorous calibration procedures that take into account the structure of financial energy data, in particular,  the fact that data is not available on weekends and holidays, making the data frequency or, equivalently, the data step uneven. The possibility of missing data only exacerbates this effect. In this context, the estimation of mean-reversion rates in the models considered becomes non-trivial and we fully address this issue.
We also take into account seasonal effects that are typical for energy markets by allowing term structures of model parameters such as volatility and correlations. We provide confidence intervals for the estimates to see if they are statistically significant, which should be defined on a case-by-case basis. For some data sets, in order to increase the robustness of estimators, the granularity of the model term structures should be chosen in such a way that it leads to smaller confidence intervals, for example, the correlation term structure can be chosen to be flat and the volatility term structure can be chosen to be seasonal rather than monthly. These are the modelling choices that affect the value of calibrated parameters and the robustness of estimators.

The structure of the paper is as follows. First, we describe the one- and two-factor models we use for valuation and their calibration. Then we describe how these models are joint together for the multi-asset case valuation and provide calibration results of natural gas prices at several location for the US market. We also provide an example of calibration with regional ERCOT power prices.

{\bf Acknowledgements}
We would like to thank Angelo Barbieri, Tsvetan Stoyanov, and Maksim Oks at MSCI Research Valuation and Energy group for numerous fruitful discussions and comments.

\section{One-Factor Mean-Reversion Model}

The one-factor log-normal mean-reversion model is defined by the following stochastic differential equation (SDE)
\begin{equation}
\label{MRSDE}
d\log{S_t}=(\theta_t-a\log{S_t})dt+\sigma_t dW_t,
\end{equation}
where $\theta_t$ is derived from the no-arbitrage condition 
$\E_0[S_t]=F(0,t)$ under the risk-neutral measure.

Analogously, the one-factor normal mean-reversion model is defined by
the following SDE
\begin{equation}
dS_t=(\theta_t-aS_t)dt+\sigma_t dW_t
\end{equation}
with the same no-arbitrage condition.

Further on, we will focus on the log-normal mean-reversion model, the results for the normal model are very similar and can be easily obtained in the same way.

\subsection{Risk-Neutral Valuation}
\label{RNValMR}

We denote $X_t=\log{S_t}$.

It is easy to get
\begin{equation}
X_t=e^{-a(t-s)}X_s+\int_s^t e^{-a(t-u)}\theta_u du+\int_s^t e^{-a(t-u)}\sigma_u dW_u.
\end{equation}

From the no-arbitrage condition we have
\begin{equation}
\int_0^t e^{-a(t-u)}\theta_u du=\log{F(0,t)}-e^{-at}\log{F(0,0)}-
\frac{1}{2}\int_0^t e^{-2a(t-u)}\sigma_u^2 du.
\end{equation}

It follows that
\begin{equation}
S_t=F(0,t)\exp{\left(-\frac{1}{2}V^2(0,t)+\int_0^t e^{-a(t-u)}\sigma_u dW_u\right)}
\end{equation}
with
\begin{equation}
V^2(0,t)=\int_0^t e^{-2a(t-u)}\sigma_u^2 du.
\end{equation}

If we define the evolution of the forward curve to be 
$F(t,T)=\E_t[S_T]$, we get
\begin{equation}
F(t,T)=F(0,T)
\exp\left(\frac{1}{2}\bigl(1-e^{-a(T-t)}\bigr)e^{-a(T-t)}V^2(0,t)
\right)
\left(\frac{S_t}{F(0,t)}\right)^{e^{-a(T-t)}}.
\end{equation}
Thus, we see that the dynamics of the whole forward curve is explicitly defined via the dynamics of the spot.

\subsection{Calibration}

We use historical data for calibration of the model parameters. Now we are considering the model in the real-world measure, hence, the mean-reversion level function $\theta_t$ contains the market price of risk and is different from the result obtained in the previous section \ref{RNValMR}.

Different granularity can be assumed for the term structures of mean-reversion level and volatility. Without lack of generality, let us assume that the granularity of both term structures is monthly, although, our results generalize to an arbitrary time indexation scheme.

In addition, we assume yearly periodicity of volatilities, i.e. we assume that volatilities in the same calendar months in different year are the same. On one hand, this assumption allows us to capture seasonality effects, on the other hand, it enables us to use several years of data for volatility estimates making them more robust.

Thus, the term structure of volatilities is parameterized by 12 parameters $\sigma_m$, with $m=1,..,12$, and the term structure of mean-reversion level $\theta_{(m,y)}$ is parametrized by a pair of parameters $(m,y)$,  with $m=1,..,12$ and $y$ ranging over the years specified in the data set, e.g. $y=2008$, $2009$, and $2010$.

Note that we are considering data sets with varying time steps since data is not available on weekends and holidays.

Let us denote $X_t=\log{S_t}$ for log-normal and $X_t=S_t$ for normal model. Solving the above SDEs explicitly, we obtain
\begin{equation}
X_{t+dt}=e^{-adt}X_t+\theta_t\frac{1-e^{-adt}}{a}+\sigma_t\sqrt{\frac{1-e^{-2adt}}{2a}}\epsilon_t,
\end{equation}
where $\epsilon_t$ is a standard normal variable.

Let us denote 
\begin{eqnarray}
\label{ekg}
\eta_t&=&e^{-adt},\nonumber \\ 
\kappa_t&=&\frac{1-e^{-adt}}{a},\\ 
\gamma_t&=&\sqrt{\frac{1-e^{-2adt}}{2a}}.\nonumber
\end{eqnarray}

The conditional probability of going from $X_t$ to $X_{t+dt}$ is given by
\begin{equation}
\rho(X_{t+dt}|X_t, a, \sigma_t, \theta_t)=\frac{1}{\sqrt{2\pi} \sigma_t\gamma_t}\exp{\left(-\frac{(X_{t+dt}-\eta_t X_t-\theta_t\kappa_t)^2}{2\sigma_t^2\gamma_t^2}\right)}
\end{equation}

The Maximum Likelihood Function is given by
\begin{equation}
{\cal L}(a, \sigma, \theta)=\sum_{k=1}^{N-1}\log{\rho(X_{t_{k+1}}|X_{t_k}, a, \sigma_{t_k}, \theta_{t_k})},
\end{equation}
where $N$ is the number of data in the provided time series for the underlying.

Differentiating w.r.t. $\theta_{(m,y)}$, we obtain
\begin{equation}
\theta_{(m,y)}=\frac{\sum_{t_{k-1}\in (m,y)} \left(X_{t_k}-
\eta_{t_{k-1}} X_{t_{k-1}}\right)}{\sum_{t_{k-1}\in (m,y)} \kappa_{t_{k-1}}}.
\end{equation}

Thus, we see that the estimator of $\theta_t$ depends only on the mean-reversion rate parameter.

Differentiating w.r.t. $\sigma_m$, we obtain
\begin{equation}
\sigma_m=\frac{1}{N(m)}\sum_{t_{k-1}\in (m)} 
\left(\frac{X_{t_k}-\eta_{t_{k-1}} X_{t_{k-1}}-\theta_{(m,y)}(a)
\kappa_{t_{k-1}}}{\gamma_{t_{k-1}}}\right)^2.
\end{equation}
Therefore, we see that the estimator of $\sigma_t$ also depends only on the mean-reversion rate parameter.

Thus, we have proven the following result.
\begin{theorem}
The Maximum Likelihood estimators of mean-reversion level and volatility term structures for the one-factor mean-reversion model are given by

\begin{equation}
\theta_{(m,y)}(a)=\frac{\sum_{t_{k-1}\in (m,y)} \left(X_{t_k}-
\eta_{t_{k-1}} X_{t_{k-1}}\right)}{\sum_{t_{k-1}\in (m,y)} \kappa_{t_{k-1}}}
\end{equation}

and

\begin{equation}
\sigma_m(a)=\frac{1}{N(m)}\sum_{t_{k-1}\in (m)} 
\left(\frac{X_{t_k}-\eta_{t_{k-1}} X_{t_{k-1}}-\theta_{(m,y)}(a)
\kappa_{t_{k-1}}}{\gamma_{t_{k-1}}}\right)^2.
\end{equation}

The mean-reversion rate can be found by finding the global maximum of the function
\begin{equation}
{\cal L}(a)=\sum_{k=1}^{N-1}\log{\rho(X_{t_{k+1}}|X_{t_k}, a, \sigma_{t_k}(a), \theta_{t_k}(a))}.
\end{equation}

or 

\begin{equation}
{\cal L}(a)=\sum_{k=1}^{N-1} -\log{(\sigma_{t_k}(a)\gamma_{t_k})}-
\frac{\left(X_{t_{k+1}}-\eta_{t_k} X_{t_k}-\theta_{t_k}(a)\kappa_{t_k}\right)^2}{2\sigma_{t_k}^2(a)\gamma_{t_k}^2},
\end{equation}
where parameters $\gamma$ and $\kappa$ are defined by (\ref{ekg}) and also depend on $a$.

\end{theorem}

The maximum of this function can be found by using the Brent one-dimensional search method with the initial guess provided by the estimate of the mean-reversion rate based on the assumption of constant time steps (or, equivalently, on the assumption that the regression coefficients are slowly varying with time).

We note that, as usual, the unbiased variance estimator for volatility should be slighlty adjusted

\begin{equation}
\sigma_m(a)=\frac{1}{N(m)-1}\sum_{t_{k-1}\in (m)} 
\left(\frac{X_{t_k}-\eta_{t_{k-1}} X_{t_{k-1}}-\theta_{(m,y)}(a)
\kappa_{t_{k-1}}}{\gamma_{t_{k-1}}}\right)^2.
\end{equation}

\subsubsection{Constant Time Step}

Let us see how the results can be simplified if time steps are constant. 

In this case, we have
$$
\theta_{(m,y)}=\frac{1}{N(m,y)}\sum_{t_{k-1}\in (m,y)} X_{t_k}-
\frac{e^{-adt}}{N(m,y)}\sum_{t_{k-1}\in (m,y)} X_{t_{k-1}}.
$$

We introduce the mean level function $f_t$ such that
\begin{equation}
\theta_t=\partial_t \log f_t+a\log f_t.
\end{equation}
Then, it follows that
$$
f_t=\left(\prod_{t_{k-1}\in (m,y)} S_{t_{k-1}}\right)^{1/N(m,y)}
$$
and
$$
f_{t+dt}=\left(\prod_{t_{k-1}\in (m,y)} S_{t_k}\right)^{1/N(m,y)}
$$ 
for $t\in (m,y)$, i.e. the mean level is the geometric average of prices in the month $(m,y)$.

We can rewrite the model process (\ref{MRSDE}) as
\begin{equation}
d\log(S_t/f_t)=-a\log(S_t/f_t)dt+\sigma_t dW_t.
\end{equation}
Denoting $x_t=\log(S_t/f_t)$ and integrating we obtain
\begin{equation}
x_{t+dt}=\kappa x_t+\sigma_t\gamma\epsilon_t, 
\end{equation}
where $\kappa$ and $\gamma$ are constant.
We bucket the time series into calendar months over which $\sigma_t$ is also constant $\sigma_m$. Let us also denote $N(m)$ by $n$.
Then, we perform the regression to obtain
\begin{equation}
\kappa=\frac{nS_{xy}-S_xS_y}{\Delta},
\end{equation}
where
\begin{eqnarray}
S_x&=&\sum_{t_{k-1}\in (m)} x_{t_{k-1}},\nonumber\\
S_y&=&\sum_{t_{k-1}\in (m)} x_{t_k},\nonumber\\
S_{xx}&=&\sum_{t_{k-1}\in (m)} x_{t_{k-1}}^2,\\
S_{xy}&=&\sum_{t_{k-1}\in (m)} x_{t_{k-1}}x_{t_k},\nonumber\\
\Delta&=&nS_{xx}-S_x^2\nonumber
\end{eqnarray}
with the standard error given by
\begin{equation}
\sigma_\kappa^2=S_{xx}/\Delta.
\end{equation}

It follows that the mean-reversion rate is given by
\begin{equation}
a=-\log(\kappa)/dt.
\end{equation} 

When the time steps are not approximately constant, the following estimate gives a result close to the MLE result. The daycount $C$ of the time series can be computed by
\begin{equation}
C=365 N\{ {\rm num\;of\;points\;in\;time\;series} \}/N\{ {\rm time\; length\;of\;the\;time\;series} \}.
\end{equation}
The estimate of the mean-reversion rate $\hat{a}$ can be obtained 
by following the above regression procedure. After that, the estimate of the mean-reversion rate should be adjusted by the day count, i.e.
\begin{equation}
a=\hat{a}C.
\end{equation}

However, we note that for data with varying time steps this simple estimation procedure can fail for some time series, e.g. we encountered the cases when it fails for the time series of weekend power prices.

\subsection{Zero Mean-Reversion Rate Limit}

In the limit when the mean-reversion rate goes to zero, all the formulae derived for the one-factor mean-reversion model are well-defined and the one-factor mean-reversion model reduces to the Black-Scholes model
\begin{equation}
\label{BS-1}
d\log{S_t}=\mu_t dt+\sigma_t dW_t,
\end{equation}
where $\mu_t=\lim_{a\rightarrow 0} \theta_t$.

The following proposition naturally follows 
\begin{proposition}
The MLE estimators of the mean-reversion level and volatility term structures in the Black-Scholes model are given by 
\begin{equation}
\label{BS-2}
\mu_{(m,y)}=\frac{\sum_{t_{k-1}\in (m,y)} \left(X_{t_k}-
X_{t_{k-1}}\right)}{\sum_{t_{k-1}\in (m,y)} dt_{k-1}}
\end{equation}
and
\begin{equation}
\label{BS-3}
\sigma_m=\frac{1}{N(m)}\sum_{t_{k-1}\in (m)} 
\frac{\left(X_{t_k}-X_{t_{k-1}}-\mu_{(m,y)} dt_{k-1}\right)^2}{dt_{k-1}}.
\end{equation}
\end{proposition}

\subsection{Calibration Procedure Steps}

Here we summarize the calibration procedure.

1. Find the estimate of mean-reversion rate $a_0$ by performing regression under the assumption of the constant data step. (Or use an arbitrary reasonable initial estimate.)

2. Find the maximum of the ML function using the one-dimensional search procedure with the found initial value of the mean-reversion $a_0$.

3. Compute the estimators for the term structures of the mean-reversion level and volatility.

4. Find the error estimates for these estimators.

5. Find the residual/model factor time series of the model.

6. Compute the statistics of the residual time series to see if the model is a good fit to the historical data. Accept or reject the modelling hypothesis based on the Jarque-Bera and/or Kolmogorov-Smirnov tests.

We describe the computation of the error estimates and statistics of the residuals in the following sections.

\subsection{Goodness of Model Fit}

After the calibration is done and all the parameters are found, we can find the residual time series of the model
\begin{equation}
\epsilon_t=\frac{X_{t+dt}-\eta_t X_t -\theta_t\kappa_t}{\sigma_t\gamma_t}.
\end{equation}
and compute the first four moments of the residual time series (see section \ref{examples} for examples). We found that a typical situation for the real energy price data is that the first three moments are very close to normal (0,1,0), however, the fourth moment is different from 0, i.e. the distribution of the residuals is leptokurtic. Thus, as expected,  the distribution of residuals for real energy data has fat tails.

We also compute the Jarque-Bera and Kolmogorov-Smirnov statistics of the residual time series to have a more rigorous tests of normality of residuals and the goodness of fit of the model.

\subsubsection{Jarque-Bera statistics}

The Jarque-Bera statistics of a time series $x_i$ with $n$ data points is given by
\begin{equation}
JB=\frac{n}{6}(S^2+\frac{1}{4}K^2)
\end{equation}
where $S$ is the sample skewness and $K$ is the sample kurtosis
\begin{eqnarray}
S=\frac{\mu_3}{\mu_2^3},\nonumber\\ 
K=\frac{\mu_4}{\mu_2^4}-3.
\end{eqnarray}
Here, $\mu_k$ is the $k$th  moment of the time series
\begin{equation}
\mu_k=\frac{1}{n}\sum_{i=1}^n (x_i-\bar{x})^k,
\end{equation}
where $\bar{x}$ is the sample mean
\begin{equation}
\bar{x}=\frac{1}{n}\sum_{i=1}^n x_i.
\end{equation}

It is well-known that the JB statistics have asymptotically chi-square distribution with two degress of freedom, however, this approximation is good only for large sample sizes ($>2000$). Therefore, for typical sizes of financial time series, the results of Monte Carlo simulations should be used. We used the implementation of the Jarque-Bera test provided in Alglib project, which is freely available online under a general GNU license.

\subsubsection{Kolmogorov-Smirnov statistics}
We use a simple Kolmogorov-Smirnov test. First we compute the Kolmogorov-Smirnov statistics of the time series $x_i$ with $n$ data points
\begin{equation}
D_0=\max_{i=0,..,n-1}(|i/n-N(x_i)|, |(i+1)/n-N(x_i)|).
\end{equation}
Then, the p-value is given by, for example, see \cite{NR}
\begin{equation}
p={\rm Prob}(D>D_0)=Q_{KS}\left((\sqrt{n}+0.12+0.11/\sqrt{n})D_0\right)
\end{equation}
The following Kolmogorov-Smirnov test with $\alpha$ confidence level naturally follows. If $D_0>D_\alpha$, where 
$D_\alpha=Q_{KS}^{-1}(1-\alpha)$, then the null hypothesis that the distribution of residuals is normal, can be rejected with $1-\alpha$ confidence.

\section{Confidence Intervals for the Volatility, Correlation, and Mean-Reversion Rate Estimates}

We use well-known results for estimation of the confidence intervals for volatility and correlation estimates, for example, see \cite{Eydel} and \cite{Haug}.

\subsection{Confidence Intervals for the Volatility Estimates}

Let us simplify the notation and denote $N(m)$ by $n$ and $\sigma_m$ by $\sigma$. Then
\begin{eqnarray}
P\left[\sigma_{lb}\leq\sigma\leq\sigma_{ub}\right]=1-\alpha,\nonumber\\
\sigma_{lb}=\sigma \sqrt{\frac{n-1}{\chi^2_{(n-1;\alpha/2)}}},\\
\sigma_{ub}=\sigma \sqrt{\frac{n-1}{\chi^2_{(n-1;1-\alpha/2)}}},\nonumber
\end{eqnarray}
where $\alpha$ is the confidence level and $\chi^2_{(n-1;\alpha/2)}$ is the value of the chi-square distribution with $n-1$ degrees of freedom and a confidence level $1-\alpha$.

\subsection{Confidence Intervals for the Correlation Estimates}

Let us simplify the notation and denote $N(m)$ by $n$ and $\rho_m$ by $\rho$. Then, using Fisher's transformation, we have the following bounds on the correlation estimates
\begin{eqnarray}
P\left[\rho_{lb}\leq\rho\leq\rho_{ub}\right]=1-\alpha,\nonumber\\
\rho_{lb}=\frac{\exp{(2z_{lb})}-1}{\exp{(2z_{lb})}+1},\\
\rho_{ub}=\frac{1-\exp{(-2z_{lb})}}{1+\exp{(-2z_{lb})}},\nonumber
\end{eqnarray}
where
\begin{eqnarray}
z_{lb}=z-c/\sqrt{n-3},\nonumber\\
z_{ub}=z+c/\sqrt{n-3}
\end{eqnarray}
with
\begin{eqnarray}
z=\frac{1}{2}\log{\left(\frac{1+\rho}{1-\rho}\right)},\nonumber\\
c=N^{-1}(1-\alpha/2).
\end{eqnarray}
$N^{-1}$ is the inverse of the cumulative normal distribution and $\alpha$ is the confidence level.

\subsection{Confidence Interval for the Mean-Reversion Estimate}

A standard error estimate $\sigma_{a_0}$ for the initial value of the mean-reversion rate $a_0$ follows from the regression procedure. A crude standard error estimate of the MLE estimator of the mean-reversion rate can be obtained in a simple way by scaling the standard error estimate coming from regression by the ratio of the MLE and initial value of the mean-reversion rate:
$$
\sigma_a=\frac{a}{a_0}\sigma_{a_0}.
$$

In order to estimate $\sigma_a$ more rigorously, Fisher information should be employed.
\section{Two-Factor Spot-Prompt Model}

The two-factor Spot-Prompt model is a natural extension of the one-factor log-normal mean-reversion model which uses the prompt-month forward price (the "index") as the stochastic mean-reversion level, see \cite{TS}.
The model is defined by the following system of SDEs 
\begin{eqnarray}
d\log{S_t}&=&(\theta_t+a\log{I_t}-a\log{S_t})dt+\sigma_t^S dW_t,
\nonumber \\
d I_t&=&I_t\sigma_t^I dB_t,\\
\Covar [dB_t, dW_t] &=&\rho_t dt,\nonumber
\end{eqnarray}
where $\theta_t$ is derived in the risk-neutral measure from the series of no-arbitrage conditions described in the next section.  

\subsection{Risk-Neutral Valuation}

The Spot-Prompt model is a two-factor model, where the evolution of the forward curve is described by the rolling prompt contract. Inside each month, the expected value of the spot is given by the level of the just expired forward contract (the index). The spot mean-reverts to the stochastic level of the prompt contract. The prompt contract follows the driftless geometric Brownian motion.

Let us denote by $T_i$ the end of each month from the value date $t=0$ to the expiry of the contract $T$, i.e. $T_0\leq 0\leq T_1<..<T_n\leq T\leq T_{n+1}$(following NYMEX conventions, $T_i$ falls on a business day three business days prior to the first delivery date). Let us assume that the forward curve is given by specifying the prices of forward contracts expiring at the end of each month $F(0, T_i)$. We assume that the dynamics of forward contracts is given by the Black model:
\begin{equation}
\label{SP-1}
dF(t, T_i)=F(t,T_i)\sigma_F(t,T_i)dB_i
\end{equation}
for $t < T_i$. We make the following two assumptions that will allow us to describe the evolution of the forward curve using one-factor model of the rolling prompt:

1. All forward contracts are perfectly correlated,

2. The local volatility function for each forward contract is given by 
$$
\sigma_F(t, T_i)=\exp(-b(T_i-t))\sigma_F(T_i),
$$
where $b$ is a forward mean-reversion rate. Hence, we include the time-to-maturity/Samuelson effect through the introduction of the forward mean-reversion rate.

Let $I_t=F(t,T_i)$ for $T_{i-1}\leq t < T_i$ denote the prompt contract at time $t$ and $\sigma_I(t)=\sigma_F(t, T_i)$ for 
$T_{i-1}\leq t < T_i$ denote the term structure of prompt contract volatility, then, under the above assumptions, the dynamics of the forward curve is effectively given by the dynamics of the rolling prompt:
\begin{equation}
\label{SP-2}
dI_t=I_t\sigma_I(t)dB_t.
\end{equation}

Note that the process $I_t$ is discontinuous at $T_i$ since at every $T_i$ the value is switched from $F(T_i,T_i)$ at $T^-_{i}$ to $F(T_i,T_{i+1})$ at $T^+_i$.

In the Spot-Prompt model, we assume that the following mean-reverting process describes the risk-neutral dynamics of the spot:
\begin{equation}
\label{SP-3}
d\log{S_t}=(\theta_t+a\log{I_t}-a\log{S_t})dt+\sigma_t^S dW_t,
\end{equation}
where $\Covar [dB_t, dW_t]=\rho_t dt$, $\sigma_S(t)$ is the deterministic term structure of spot volatility, and $\theta_t$ is the mean reversion level to be derived from the following no-arbitrage assumption:
\begin{equation}
\label{SP-4}
\E_{T_i}[S_t]=I_{T^-_i}=F(T_i,T_i)
\end{equation}
for $T_i\leq t < T_{i+1}$. This assumption simply means that there is no riskless profit to be made by buying index and selling spot or vice versa, also see \cite{Eydel}.

We note that the initial no-arbitrage condition $\E_0[S_t]=F(0,t)$ holds by the tower law property and the fact the index process is a martingale.

In order to find the level $\theta(t)$, we derive the following equality (we omit the details of this derivation):
\begin{eqnarray}
\label{SP-5}
\int_s^t e^{-a(t-u)}\theta(u)du=
\log{\left(\frac{\E_{T_i}[S_t]}{\E_{T_i}[I_t]}\right)}-
e^{-a(t-s)}\log{\left(\frac{\E_{T_i}[S_s]}{\E_{T_i}[I_s]}\right)}-\nonumber\\
\frac{V(T_i,t)}{2}+e^{-a(t-s)}\frac{V(T_i,s)}{2}+
\frac{1}{2}\int_s^t e^{-a(t-u)}\sigma^2_I(u)du-\nonumber\\
\int_s^t e^{-a(t-u)}\sigma_I(u)\sigma_S(u)\rho(u)du
\end{eqnarray}
for $T_i\leq s < t < T_{i+1}$, where
\begin{equation}
V(s,t)=\Var_s[\log{\left(\frac{S_t}{I_t}\right)}]
\end{equation}
and $\Var_s$ is the conditional variance at time $t$ given the information at time $s$ with $s < t$. Computing $V(s,t)$ explicitly, we 
have
\begin{equation}
\label{SP-6}
V(s,t)=\int_s^t \sigma^2(u)du
\end{equation}
with
\begin{equation}
\label{SP-7}
\sigma^2(u)=\sigma_S^2(u)+\sigma_I^2(u)-2\sigma_S(u)\sigma_I(u)\rho(u)
\end{equation}

We note that (\ref{SP-5}) follows from the dynamics of the underlyings (\ref{SP-2}) and (\ref{SP-3}) and is true for any no-arbitrage assumptions.
From (\ref{SP-5}) it follows that
\begin{equation}
\label{SP-8}
\theta(t)=\log{\left(\frac{\E_{T_i}[S_t]}{\E_{T_i}[I_t]}\right)}+
\frac{1}{2}V(T_i,t)-\frac{\sigma_S^2(t)}{2a}
\end{equation}
for $T_i\leq t < T_{i+1}$.
From the dynamics of the rolling prompt, we compute
\begin{equation}
\label{SP-9}
\E_{T_i}[I_t]=I_{T_i}
\end{equation}
for $T_i\leq t < T_{i+1}$. Therefore, from (\ref{SP-4}) and (\ref{SP-9}) it follows that
\begin{equation}
\label{SP-10}
\theta(t)=\frac{1}{2}V(T_i,t)-\frac{\sigma_S^2(t)}{2a}
\end{equation}
for $T_i\leq t < T_{i+1}$.

From (\ref{SP-10}) we see that the Spot-Prompt model with the imposed no-arbitrage assumption (\ref{SP-4}) can be easily implemented in the Monte Carlo simulation framework, see \cite{FEAST}.

$S_t$ is a lognormal variable. For the simulation, we use the expected value of $\log{S_t}$
\begin{eqnarray}
\label{SP-11}
\E_s[\log{S_t}]=e^{-a(t-s)}\log{S_s}+(1-e^{-a(t-s)})\log{I_s}-\\
\frac{1}{2}V(T_i,t)+\frac{1}{2}e^{-a(t-s)}V(T_i,s)+\frac{1}{2}V(s,t)-\frac{1}{2}\Var_s[\log{S_t}]
\end{eqnarray}
and the variance of $\log{S_t}$
\begin{eqnarray}
\label{SP-12}
&&\Var_s[\log{S_t}]=\int_s^t\sigma_I^2(u)du-\nonumber\\
&&2\int_s^t e^{-a(t-s)}(\sigma_I^2(u)-\sigma_I(u)\sigma_S(u)\rho(u))du+
V(s,t)
\end{eqnarray}
for $T_i\leq s < t < T_{i+1}$.
We note that the spot process is continuous inside each month but is discontinuous at $T_i$ in general (compare with \cite{Ronn}).

We also note that by the tower law and the fact that the index is a martingale it follows that
\begin{equation}
\E_s[\log{S_t}]=F(s,T_i)
\end{equation}
for $s<T_i\leq t < T_{i+1}$.

\subsection{Calibration}

In the real-world measure the model can be written as
\begin{eqnarray*}
d\log{S_t}&=&(\theta_t+a\log{I_t}-a\log{S_t})dt+\sigma_t^S dW_t,\\
d\log{I_t}&=&\mu_t dt+\sigma_t^I dB_t,\\
\Covar [dB_t, dW_t]&=&\rho_t dt,
\end{eqnarray*}

We can rewrite it as
\begin{eqnarray}
d\log{(S_t/I_t)}&=&(\tilde{\theta}_t-a\log{(S_t/I_t)})dt+\sigma_t d\tilde{W}_t,\nonumber \\
d\log{I_t}&=&\mu_t dt+\sigma_t^I dB_t,\\
\Covar [dB_t, d\tilde{W}_t]&=&\tilde{\rho}_t dt,\nonumber
\end{eqnarray}
where $d\tilde{W}_t=(\sigma_t^SdW_t-\sigma_t^IdB_t)/\sigma_t$ and, therefore, 
\begin{eqnarray}
\label{sigma}
\sigma_t^2&=&(\sigma_t^S)^2+(\sigma_t^I)^2-2\sigma_t^S\sigma_t^I\rho_t,\nonumber \\
\tilde{\rho}_t&=&(\sigma_t^S\rho_t-\sigma_t^I)/\sigma_t,\\
\tilde{\theta}_t&=&\theta_t-\mu_t.\nonumber
\end{eqnarray}

It follows that effectively we have one-factor mean-reversion model on the quotient of $S_t$ and $I_t$. Thus we can apply the results of the previous chapter to get the mean-reversion rate $a$, the term structure of $\tilde{\theta}_t$ and $\sigma_t$.

We also apply the results of the section on the Black-Scholes model (\ref{BS-2}) and (\ref{BS-3}) to get the term structure of $\mu_t$ and $\sigma_t^I$.

Thus, we have the following result.

\begin{proposition}
The mean-reversion rate estimate for the Spot-Prompt model follows from the MLE estimate for the one-factor mean-reversion model that describe the dynamics of the quotient process of spot $S_t$ and index $I_t$.
\end{proposition}

Next, we proceed to finding the term structure correlation between the model spot and prompt factors and the term structure of spot volatilities.
 
Denoting $X_t=\log(S_t/I_t)$ and $Y_t=\log(I_t)$, we have
\begin{eqnarray}
\label{SPMSol}
X_{t+dt}=\eta_t X_t+\tilde{\theta}\kappa_t+
\sigma_t\gamma_t\epsilon_t,\nonumber\\
Y_{t+dt}=Y_t+\mu_t dt+\sigma_t^I\sqrt{dt}\xi_t,
\end{eqnarray}

Let us introduce the following time series:
\begin{eqnarray}
\tilde{X_t}&=&(X_{t+dt}-e^{-adt}X_t-\theta_t\eta_t)/\eta_t,\nonumber\\
\tilde{Y_t}&=&Y_{t+dt}-Y_t-\mu_t dt.
\end{eqnarray}
Then, the covariance between these time series is given by
\begin{equation}
\Covar[\tilde{X_t}, \tilde{Y_t}]=\sigma_t^I(\sigma_t^S\rho_t-\sigma_t^I).
\end{equation}
From here, since we know $\sigma_t^I$, we find $\sigma_t^S\rho_t$.
Then, plugging it in (\ref{sigma}), we find $\sigma_t^S$,
$$
\sigma_t^S=\sqrt{\sigma_t^2-(\sigma_t^I)^2+2\sigma_t^S\sigma_t^I\rho_t}.
$$
Finally, dividing the product $\sigma_t^S\rho_t$ by $\sigma_S$, we get $\rho_t$.

\subsubsection{Spot Factor Time Series}

Rewriting (\ref{SPMSol}), we have
\begin{eqnarray}
X_{t+dt}&=&\eta_t X_t+\tilde{\theta}\kappa_t+\sigma_t^S\gamma_t\epsilon_t^S-\sigma_t^I\gamma_t\nu_t,\nonumber\\
Y_{t+dt}&=&Y_t+\mu_t dt+\sigma_t^I\sqrt{dt}\xi_t,
\end{eqnarray}
where $\epsilon_t^S$, $\nu_t$, and $\xi_t$ are correlated standard normal variables. We need to find time times series $\epsilon_t^S$ for the computation of model factor correlations when the two-factor model is included in the joint multi-asset model.
 
First, we compute the correlation between $\nu_t$ and $\xi_t$
\begin{equation}
\rho(\nu_t,\xi_t)=\frac{\kappa_t}{\gamma_t\sqrt{dt}}=\sqrt{\frac{2(1-e^{-adt})}{adt(1+e^{-adt})}}.
\end{equation}
We notice that the function on the right is very close to identity even when the mean-reversion rates are very high, e.g. when $adt$ is $0.5$, the correlation is $0.99$, when $adt$ is $1$, the correlation is $0.96$.
Thus, we can approximate $\nu_t$ by $\xi_t$, to have
\begin{equation}
\epsilon_t^S \approx \frac{1}{\sigma_t^S\gamma_t}\left(X_{t+dt}-\eta_t X_t-\tilde{\theta_t}\kappa_t+\frac{\gamma_t}{\sqrt{dt}}(Y_{t+dt}-Y_t-\mu_t dt)\right).
\end{equation}
We ran tests to see that this approximation works quite well, which follows from the statistics of the residual time series for different data sets.

\section{The Joint Model of Several Underlyings}

Here we describe the joint model that we use for the description of the joint dynamics of several underlyings. Let us assume that we need to calibrate $n$ underlyings. Without loss of generality, we assume that the first $k$ underlyings are described by the Spot-Prompt model and the rest -- by the log-normal mean-reversion model. Then, we have
\begin{equation}
d\log{S_i(t)}=(\theta_i(t)+a_i\log{I_i(t)}\delta_{i}-a_i\log{S_i(t)})dt+\sigma_{S_i}(t)dW_i(t),
\end{equation}
with $\delta_i=1$ for $i=1,..,k$ and $\delta_i=0$ for $i=k+1,..,n$, 
and for the index
\begin{equation}
d\log{I_i(t)}=\mu_i(t)dt+\sigma_{I_i}(t) dB_i(t),
\end{equation}
for $i=1,..,k$.
There is a correlation between factors
\begin{eqnarray}
\Covar [dW_i, dW_j]&=&\rho_{ij}(t)dt,\nonumber\\
\Covar [dW_i, dB_j]&=&q_{ij}(t)dt.
\end{eqnarray}
We note that the chosen granularity of volatility $\sigma_{S_i}$, $\sigma_{I_i}$ and correlation $\rho_{ij}(t)$, $q_{ij}(t)$  term structures as well as the granularity of the mean-reversion level $\theta_i(t)$ and drift $\mu_i(t)$ term structures is a modelling assumption that directly affects calibration results as well as robustness of estimates. It should be chosen on a case by case basis depending on the data.

In order to calibrate the joint model, we first calibrate each model separately to get $a_i$, the term structures $\sigma_{S_i}(t)$, and $\sigma_{I_i}(t)$, and $\theta_i(t)$ and $\mu_i(t)$. After that we obtain the normalized time series of model factors corresponding to $dW_i(t)$ and $dB_i(t)$ driving factors. Then we compute the correlations between these factors to get the term structures of correlations $\rho_{ij}(t)$ and $q_{ij}(t)$.

We run the Jarque-Bera and Kolmogorov-Smirnov tests on these time series to define how good of a fit the model is for the provided data.
In order to improve the fit, we can introduce a simple procedure for removing outliers. We take the distribution of the residuals of each model and throw away some percentage of outliers, e.g. from 1\% to 5\% percent in the tails. We can think of these points as points corresponding to jumps that were not accounted by us in the model. We saw in our examples, that this procedure improves the fit of the model to the data.

We estimate the confidence intervals for the estimates in order to see if the modelling assumption on the granularities of term structures is good for provided data. The correlation estimates notoriously have big estimation errors, so the confidence intervals for correlations should be carefully checked. If the confidence intervals are too big and cannot be accepted the coarser granularity should be assumed on the term structure and the model should be recalibrated.

\section{Calibration Examples}
\label{examples}

We use daily closing prices of spot natural gas at several location. In the first example we use around 10 years of data from 01/01/1998 to 11/25/2009 and two US location STX, in the Gulf, and M3, in the Northeast, with the log-normal mean-reversion model. 

In the second example, we use about 2 years of data from 2/13/2008 to 11/25/2009 at two US locations STX and WLA together with the index data at those locations with the spot-prompt model.

In the third example, we use regional ERCOT hourly power prices and system load from 7/1/2004 to 2/28/2010 with the log-normal mean-reversion model. 

\subsection{Log-Normal Mean-Reversion Model Daily Natural Gas Data}

Here we use around 10 years of data from 01/01/1998 to 11/25/2009 for spot natural gas prices at STX and M3 with the log-normal mean-reversion model. The data step is daily. We assume that the granularity of the term structures of mean-reversion level, volatility and correlation is monthly.

The results of the calibration are summarized below.
The mean-reversion rate estimates and their standard errors are given by
\begin{center}
\begin{tabular}{|l|l|l|l|l|l|l|}
\hline
 & STX & M3 \\ 
\hline
mr & 38.73 & 47.48 \\ 
\hline
mre & 2.59 & 2.5 \\ 
\hline
\end{tabular}
\end{center}

We assume the step following interpolation of the term structures.
The local volatility term structures are given by
\begin{center}
\begin{tabular}{|l|l|l|l|l|l|l|}
\hline
 & STX & M3 \\ 
\hline
Nov &	1.32  & 1.36 \\
\hline
Dec &	1.10 & 1.63 \\
\hline
Jan &	0.73 & 3.05 \\
\hline
Feb &	1.24 & 2.15 \\
\hline	   
Ma &	0.60 & 1.47 \\
\hline
Apr &	0.46 & 0.51 \\
\hline	   
May &	0.49 & 0.49 \\
\hline	   
Jun &	0.56 & 0.59 \\
\hline
Jul &	0.54 & 0.63 \\
\hline	   
Aug &	0.66 & 0.70 \\
\hline	   
Sep &	0.86 & 0.83 \\
\hline	   
Oct &	1.07 & 1.06 \\
\hline
\end{tabular}
\end{center}

The lower bounds of the confidence intervals are

\begin{center}
\begin{tabular}{|l|l|l|l|l|l|l|}
\hline
 & STX & M3 \\ 
\hline
Nov &	1.23 & 1.27 \\
\hline
Dec &	1.02 & 1.52 \\
\hline
Jan &	0.68 & 2.85 \\
\hline
Feb &	1.15 & 2.00 \\
\hline	   
Mar &	0.56 & 1.37 \\
\hline
Apr &	0.43 & 0.47 \\
\hline	   
May &	0.46 & 0.46 \\
\hline	   
Jun &	0.52 & 0.55 \\
\hline
Jul &	0.51 & 0.59 \\
\hline	   
Aug &	0.62 & 0.66 \\
\hline	   
Sep &	0.80 & 0.77 \\
\hline	   
Oct &	1.00 & 0.99 \\
\hline
\end{tabular}
\end{center}

The upper bounds of the confidence intervals are

\begin{center}
\begin{tabular}{|l|l|l|l|l|l|l|}
\hline
 & STX & M3 \\ 
\hline
Nov &	1.42 & 1.47 \\
\hline
Dec &	1.19 & 1.77 \\
\hline
Jan &	0.79 & 3.29 \\
\hline
Feb &	1.34 & 2.32 \\
\hline	   
Mar &	0.64 & 1.58 \\
\hline
Apr &	0.50 & 0.55 \\
\hline	   
May &	0.53 & 0.53 \\
\hline	   
Jun &	0.60 & 0.63 \\
\hline
Jul &	0.59 & 0.68 \\
\hline	   
Aug &	0.72 & 0.76 \\
\hline	   
Sep &	0.93 & 0.90 \\
\hline	   
Oct &	1.16 & 1.15 \\
\hline
\end{tabular}
\end{center}

The statistics for the residual time series are
\begin{center}
\begin{tabular}{|l|l|l|l|l|l|l|}
\hline
 & mean & stddev & skewness & kurtosis & JB stats & KS stats \\
\hline
STX & 0.0000 & 0.9987 &	0.2997 & 8.8683 & 14306.7398 & 0.1070 \\
\hline
M3 & 0.0000	& 0.9987 & 0.8369	& 12.9883 &	31055.4814 & 0.1213 \\
\hline
\end{tabular}
\end{center}
We see that the model fit is quite poor, which was expected with this long history of data.
 
In the following table, we provide the correlation estimate and the corresponding lower and upper bounds of the confidence interval estimate:
\begin{center}
\begin{tabular}{|l|l|l|l|}
\hline
& corr & lower & uppper\\
\hline
Nov & 0.9244 & 0.9077 & 0.9383 \\
\hline
Dec & 0.6724 & 0.6097 & 0.7267 \\
\hline
Jan & 0.2439 & 0.1458 & 0.3372 \\
\hline
Feb & 0.7414 & 0.6893 & 0.7859 \\
\hline
Mar & 0.5302 & 0.4529 & 0.5995 \\
\hline
Apr & 0.8464 & 0.8143 & 0.8733 \\
\hline
May & 0.9585 & 0.9493 & 0.966 \\
\hline
Jun & 0.9456 & 0.9335 & 0.9555 \\
\hline
Jul & 0.9242 & 0.9078 & 0.9378 \\
\hline
Aug & 0.9401 & 0.9271 & 0.9509 \\
\hline
Sep & 0.9396 & 0.9261 & 0.9506 \\
\hline
Oct & 0.951 & 0.9403 & 0.9599 \\
\hline
\end{tabular}
\end{center}
We see that all the estimates are robust by looking at the confidence intervals.

\subsection{Spot-Prompt Model Daily Natural Gas Data}

Here we use around 2 year of data from 2/13/2008 to 11/25/2009 at two US locations STX and WLA together with the index data at
those locations with the Spot-Prompt model. The data step is daily. We assume that the granularity of the term structures of mean-reversion level, volatility is monthly and correlation is flat.

The results of the calibration are summarized below.
The mean-reversion rate estimates and their standard errors are given by
\begin{center}
\begin{tabular}{|l|l|l|}
\hline
 & STX & WLA  \\ 
\hline
mr & 156.63 & 163.99 \\
\hline
mre & 19.79 & 20.67 \\
\hline
\end{tabular}
\end{center}
We note that the mean-reversion rate estimates are typically higher when estimated with the Spot-Prompt model than with the one-factor mean-reversion model.

We assume the step following interpolation of the term structures.
The local volatility term structures are given by
\begin{center}
\begin{tabular}{|l|l|l|l|l|}
\hline
 & STX & STXIndex & WLA & WLAIndex  \\
\hline 
Nov &	1.88 & 0.82 & 1.75 & 0.74 \\
\hline
Dec &	0.93 & 0.63 & 0.79 & 0.60 \\
\hline
Jan & 0.90 & 0.51 & 0.73 & 0.50 \\
\hline
Feb &	0.72 & 0.54 & 0.70 & 0.48 \\
\hline
Mar &	0.62 & 0.81 & 0.58 & 0.69 \\
\hline
Apr & 0.47 & 0.43 & 0.48 & 0.43 \\
\hline
May &	0.54 & 0.70 & 0.49 & 0.70 \\
\hline
Jun &	0.56 & 0.49 & 0.57 & 0.49 \\
\hline
Jul &	0.57 & 0.72 & 0.57 & 0.74 \\
\hline
Aug & 0.67 & 0.53 & 0.67 & 0.56 \\
\hline
Sep &	1.50 & 1.28 & 1.38 & 1.31 \\
\hline
Oct &	1.82 & 0.78 & 1.77 & 0.80 \\
\hline
\end{tabular}
\end{center}

The lower bounds of the confidence intervals are
\begin{center}
\begin{tabular}{|l|l|l|l|l|}
\hline
 & STX & STXIndex & WLA & WLAIndex  \\
\hline 
Nov &	1.53 & 0.66 & 1.42 & 0.60\\
\hline
Dec &	0.71 & 0.48 & 0.60 & 0.46\\
\hline
Jan &	0.68 & 0.39 & 0.56 & 0.38\\
\hline
Feb &	0.57 & 0.43	& 0.56 & 0.38\\
\hline
Mar &	0.51 & 0.66 & 0.48 & 0.57\\
\hline
Apr &	0.39 & 0.36 & 0.39 & 0.36\\
\hline
May &	0.44 & 0.57 & 0.40 & 0.58 \\
\hline
Jun &	0.46 & 0.40 & 0.47 & 0.41 \\
\hline
Jul &	0.47 & 0.60 & 0.47 & 0.61 \\
\hline
Aug &	0.55 & 0.44 & 0.55 & 0.46 \\
\hline
Sep &	1.23 & 1.05 & 1.13 & 1.08 \\
\hline
Oct & 1.51 & 0.64 & 1.46 & 0.66 \\
\hline
\end{tabular}
\end{center}

The upper bounds of the confidence intervals are
\begin{center}
\begin{tabular}{|l|l|l|l|l|}
\hline
 & STX & STXIndex & WLA & WLAIndex  \\
\hline 
Nov & 2.46 & 1.07 & 2.28 & 0.97 \\
\hline
Dec &	1.37 & 0.92 & 1.16 & 0.88 \\
\hline
Jan &	1.32 & 0.74 & 1.07 & 0.74 \\
\hline
Feb &	0.95 & 0.72 & 0.93 & 0.64 \\
\hline
Mar &	0.78 & 1.02 & 0.74 & 0.88 \\
\hline
Apr &	0.60 & 0.54 & 0.60 & 0.55\\
\hline
May &	0.69 & 0.89 & 0.63 & 0.90\\
\hline
Jun &	0.71 & 0.62 & 0.72 & 0.63\\
\hline
Jul &	0.72 & 0.91 & 0.73 & 0.93 \\
\hline
Aug &	0.85 & 0.68 & 0.86 & 0.71 \\
\hline
Sep &	1.91 & 1.63 & 1.76 & 1.67 \\
\hline
Oct &	2.30 & 0.98 & 2.23 & 1.01 \\
\hline
\end{tabular}
\end{center}

The statistics for the residual time series are given by
\begin{center}
\begin{tabular}{|l|l|l|l|l|l|l|}
\hline
 & mean & stddev & skewness & kurtosis & JB stats & KS stats \\
\hline
STX & 0.0191 & 1.0018 &	-0.4551 & 1.9719 & 88.8309 & 0.0605 \\
\hline
STXIndex & -0.0126 & 0.9876 &	0.3218 & 1.2814 &	38.7256 & 0.0503\\
\hline
WLA &	0.0146 & 1.0045 &	-0.3954 & 2.3440 & 115.2543 &	0.0565 \\
\hline
WLAIndex & -0.0151 & 0.9876 &	0.3836 & 1.5635 &	57.1260 & 0.0583\\
\hline
\end{tabular}
\end{center}
We see that the model fit is much better in this case than in the previous example.
 
The correlation matrix is given by
\begin{center}
\begin{tabular}{|l|l|l|l|}
\hline
STX & STXIndex & WLA & WLAIndex  \\
\hline 
1.00 & -0.12 & 0.94 & -0.12\\
\hline
-0.12 & 1.00 & -0.14 & 0.96 \\
\hline
0.94	& -0.14 & 1.00 & -0.15 \\
\hline
-0.12 & 0.96 & -0.15 & 1.00 \\
\hline
\end{tabular}
\end{center}
and the corresponding lower bound of the confidence interval estimate
\begin{center}
\begin{tabular}{|l|l|l|l|}
\hline
STX & STXIndex & WLA & WLAIndex  \\
\hline
1.00	& -0.21 & 0.93 & -0.21\\
\hline
-0.21 & 1.00 & -0.23 & 0.95 \\
\hline
0.93 & -0.23 & 1.00 & -0.23 \\
\hline
-0.21 & 0.95 & -0.23 & 1.00 \\
\hline
\end{tabular}
\end{center}
and the upper bound of the confidence interval estimate
\begin{center}
\begin{tabular}{|l|l|l|l|}
\hline
STX & STXIndex & WLA & WLAIndex  \\
\hline
1.00 & -0.03 & 0.95 & -0.03 \\
\hline
-0.03	& 1.00 & -0.05 & 0.96 \\
\hline
0.95 & -0.05 & 1.00 & -0.05 \\
\hline
-0.03 & 0.96 & -0.05 & 1.00 \\
\hline
\end{tabular}
\end{center}

\subsection{Log-Normal Mean-Reversion Model Hourly Load and Power Data}
Here we use around 5.5 years of hourly ERCOT power locational marginal prices (LMP) and load data from 7/1/2004 12am to 2/28/2010 11pm with the log-normal mean-reversion model. The data step is hourly. We assume that the granularity of the term structures of mean-reversion level, volatility, correlation is monthly. The valuation date is the next day after the the last date in the time series, i.e. 3/1/2010.

The results of the calibration are summarized below.
The mean-reversion rates and standard error estimates are given by
\begin{center}
\begin{tabular}{|l|l|l|}
\hline
 & Load & LMP  \\ 
\hline
mr & 476.73 & 668.39 \\
\hline
mre & 13.18 & 15.86 \\
\hline
\end{tabular}
\end{center}

We assume the step following interpolation of the term structures.
The local volatility term structures are given by
\begin{center}
\begin{tabular}{|l|l|l|}
\hline
 & Load & LMP  \\
\hline
Mar & 8.37 & 14.22\\
\hline
Apr &	9.30 & 14.15\\
\hline
May &	9.32 & 15.51\\
\hline
Jun &	8.36 & 15.17\\
\hline
Jul &	7.91 & 13.47\\
\hline
Aug &	7.96 & 13.73\\
\hline
Sep &	8.94 & 14.50\\
\hline
Oct &	9.28 & 15.89\\
\hline
Nov &	8.78 & 15.39\\
\hline
Dec &	7.73 & 14.76\\
\hline
Jan &	7.31 & 15.22\\
\hline
Feb &	7.26 & 14.19\\
\hline
\end{tabular}
\end{center}

The lower bounds of the confidence intervals are
\begin{center}
\begin{tabular}{|l|l|l|}
\hline
 & Load & LMP  \\
\hline
Mar &	8.18 & 13.91\\
\hline
Apr &	9.09 & 13.83\\
\hline
May &	9.11 & 15.17\\
\hline
Jun &	8.17 & 14.82\\
\hline
Jul &	7.75 & 13.20\\
\hline
Aug &	7.80 & 13.45\\
\hline
Sep &	8.76 & 14.20\\
\hline
Oct &	9.09 & 15.57\\
\hline
Nov &	8.60 & 15.07\\
\hline
Dec &	7.58 & 14.46\\
\hline
Jan &	7.16 & 14.91\\
\hline
Feb &	7.11 & 13.89\\
\hline
\end{tabular}
\end{center}

The upper bounds of the confidence intervals are
\begin{center}
\begin{tabular}{|l|l|l|}
\hline
 & Load & LMP  \\
\hline
Mar &	8.56 & 14.55\\
\hline
Apr &	9.52 & 14.49\\
\hline
May &	9.54 & 15.87\\
\hline
Jun &	8.55 & 15.53\\
\hline
Jul &	8.08 & 13.76\\
\hline
Aug &	8.13 & 14.02\\
\hline
Sep &	9.14 & 14.81\\
\hline
Oct &	9.47 & 16.23\\
\hline
Nov &	8.97 & 15.72\\
\hline
Dec &	7.90 & 15.08\\
\hline
Jan &	7.46 & 15.55\\
\hline
Feb &	7.42 & 14.51\\
\hline
\end{tabular}
\end{center}
We note that the mean-reversion rates and volatilities are much higher for hourly power data than for daily gas data, this is a typical feature of the results of model calibration for power data.

Here are the statistics for the residual time series
\begin{center}
\begin{tabular}{|l|l|l|l|l|l|l|}
\hline
 & mean & stddev & skewness & kurtosis & JB stats & KS stats \\
\hline
Load & 0.0000 & 0.9999 & 0.0022 & -0.0532 & 5.8915 & 0.0116 \\
\hline
LMP & 0.0000 & 0.9999 &	0.7021 & 3.1254 &	24288.7583 & 0.0673\\
\hline
\end{tabular}
\end{center}
We see that the model fit is relatively good for the Load data.
 
In the following table, we provide the correlation estimate and the corresponding lower and upper bounds of the confidence internval estimate:
\begin{center}
\begin{tabular}{|l|l|l|l|}
\hline
& corr & lower & uppper\\
\hline
Mar & 0.706 & 0.6895 & 0.7218 \\
\hline
Apr & 0.638 & 0.6182 & 0.657 \\
\hline
May & 0.6142 & 0.5938 & 0.6338 \\
\hline
Jun & 0.7176 & 0.7014 & 0.7331 \\
\hline
Jul & 0.7294 & 0.7153 & 0.7428 \\
\hline
Aug & 0.7035 & 0.6883 & 0.718 \\
\hline
Sep & 0.649 & 0.6313 & 0.6659 \\
\hline
Oct & 0.653 & 0.6359 & 0.6695 \\
\hline
Nov & 0.6968 & 0.6811 & 0.7118 \\
\hline
Dec & 0.7233 & 0.709 & 0.737 \\
\hline
Jan & 0.6971 & 0.6817 & 0.7119 \\
\hline
Feb & 0.7006 & 0.6846 & 0.716 \\
\hline
\end{tabular}
\end{center}

\section{Conclusion}

We developed a simple robust approach for the joint historical calibration of several energy underlyings. This approach takes into account seasonality effects and uneven frequency of data. It allows to choose different granularity of model parameter term structures that would provide more robust estimates based on the computed confidence intervals of model parameters. We also provided a simple way to check the goodness of model fit. It shows whether the chosen models for each underlying are a good choice for the description of the underlying dynamics. A basic procedure of how to remove data outliers was also briefly mentioned. More elaborate methods for dealing with data outliers is an interesting question and can be a topic of a future study.

We presented several examples of calibration results for several data sets. These examples provide a good illustration of typical parameter values and term structures for natural gas and power underlyings. This provides a good benchmark and guidance for calibration with other energy data sets.

\end{document}